\documentclass[symmetry,article,accept,pdftex,oneauthor]{Definitions/mdpi} 

\firstpage{1} 
\pubvolume{16}
\issuenum{1}
\articlenumber{100}
\pubyear{2024}
\copyrightyear{2024}
\externaleditor{Academic Editor: Kazuharu Bamba}
\datereceived{25 December 2023}
\daterevised{11 January 2024}
\dateaccepted{12 January 2024}
\datepublished{14 January 2024} 
\hreflink{https://doi.org/10.3390/sym16010100} 
\doinum{10.3390/sym16010100}

\Title{Time--Energy Uncertainty Relation in Nonrelativistic\newline Quantum Mechanics}

\TitleCitation{Time--Energy Uncertainty Relation in Nonrelativistic Quantum Mechanics}

\Author{Danko D. Georgiev \orcidA{}}

\AuthorNames{Danko D. Georgiev}

\AuthorCitation{{Georgiev, D.D.}
}

\address[1]{%
Department of Physics and Astronomy, University of California, Los Angeles, CA 90095, USA; danko@ucla.edu}

\abstract{The time--energy uncertainty relation in nonrelativistic quantum mechanics has been intensely debated with regard to its formal derivation, validity, and physical meaning. Here, we analyze two formal relations proposed by Mandelstam and Tamm and by Margolus and Levitin  and evaluate their validity using a minimal quantum toy model composed of a single qubit inside an external magnetic field. We show that the $\ell_1$ norm of energy coherence $\mathcal{C}$ is invariant with respect to the unitary evolution of the quantum state. Thus, the $\ell_1$ norm of energy coherence $\mathcal{C}$ of an initial quantum state is useful for the classification of the ability of quantum observables to change in time or the ability of the quantum state to evolve into an orthogonal state. In the single-qubit toy model, for quantum states with the submaximal $\ell_1$ norm of energy coherence, $\mathcal{C}<1$, the Mandelstam--Tamm and Margolus--Levitin relations generate instances of infinite ``time uncertainty'' that is devoid of physical meaning. Only for quantum states with the maximal $\ell_1$ norm of energy coherence, $\mathcal{C}=1$, the Mandelstam--Tamm and Margolus--Levitin relations avoid infinite ``time uncertainty,'' but they both reduce to a strict equality that expresses the Einstein--Planck relation between energy and frequency. The presented results elucidate the fact that the time in the Schr\"{o}dinger equation is a scalar variable that commutes with the quantum Hamiltonian and is not subject to statistical variance.}

\keyword{quantum dynamics; time; uncertainty principle}

\begin{document}


\section{Introduction}
\label{sec:1}

In the early days of the study of quantum mechanics, Werner Heisenberg~\cite{Heisenberg1927,Heisenberg1983}, Niels Bohr~\cite{Bohr1928}, Albert Einstein~\cite{Einstein1931}, Max Born, and Pascual Jordan~\cite{Born1930} considered \emph{time} and \emph{energy} as physical quantities, which \emph{{do not commute}}. The derivation of the time--energy uncertainty relation was based on thought experiments, such as Heisenberg's microscope~\cite{Heisenberg1927,Heisenberg1983} or Einstein's photon box~\cite{Bohr1983,Bohr1996}, where the goal was to employ physical intuition in order to arrive at the desired final inequality.
Despite the lack of mathematical rigor, one still had to confront the fact that   time $t$ is a real-valued \emph{{scalar}} in the Schr\"{o}dinger equation~\cite{Schrodinger1926,Schrodinger1928,Dirac1967}
\begin{equation}
\imath\hbar\frac{d}{dt}|\Psi\rangle=\hat{H}|\Psi\rangle
\label{eq:1}
\end{equation}
and commutes with the quantum Hamiltonian operator $\hat{H}$, namely
\begin{equation}
	[\hat{H}, t] = \hat{H} t - t\hat{H} = 0 .
\end{equation}
To highlight the problem, Wolfgang Pauli~\cite{Pauli1933,Pauli1980} proved as a theorem in 1933 that in nonrelativistic quantum mechanics, there cannot exist a time operator $\hat{t}$ such that
\begin{equation}
[\hat{H},\hat{t}] = \hat{H}\hat{t} - \hat{t}\hat{H} = - \imath\hbar \hat{I} .
\end{equation}

Following the discovery of the generalized Ehrenfest theorem~\cite{Ehrenfest1927} and the Robertson--Schr\"{o}dinger uncertainty relation~\cite{Robertson1929,Schrodinger1930}, different authors have attempted to use the quantum dynamics of quantum observables in order to construct time--energy uncertainty relations based on the statistical \emph{{standard deviations}} of some dynamic quantum observable and the total energy {(for comprehensive reviews, see Refs.~\cite{Busch2002,Deffner2017})}.
The motivation behind such efforts is the practical consideration that \emph{{time}} can only be measured by observing some physical quantity that is dynamically \emph{{changing}}.
In this work, we will investigate the validity and physical interpretation of  two alternative uncertainty relations proposed in 1945 by Mandelstam and Tamm~\cite{Mandelstam1945} and in 1998 by Margolus and Levitin~\cite{Margolus1998}.
We will show that the claimed time--energy uncertainty relations do not really pertain to physical time in the nonrelativistic setting, and whenever they faithfully represent time, the relations reduce to the strict Einstein--Planck  relation between energy and frequency~\cite{Bhaumik2023}.

The presentation is organized as follows:
In Section~\ref{sec:2}, we introduce the Mandelstam--Tamm uncertainty relation.
In Section~\ref{sec:3}, we derive the conservation of total energy and its quantum statistics, including the energy variance $\textrm{Var}(\hat{H})$, energy standard deviation~$\Delta \hat{H}$, and the $\ell_1$ norm of energy coherence~$\mathcal{C}$.
In Section~\ref{sec:4}, we illustrate the dynamics of a minimal quantum toy model and plot the Mandelstam--Tamm quantity~$\Delta T$ purported to measure ``time uncertainty''.
In Section~\ref{sec:5}, we introduce and analyze the Margolus--Levitin uncertainty relation based on the minimal time~$\tau_{\perp}$ taken by the quantum state vector to evolve into an orthogonal state.
Finally, in Section~\ref{sec:6}, we summarize our results on the performance of the two alternative time--energy uncertainty relations and show that their meaningful physical interpretation is just a manifestation of the Einstein--Planck relation between energy and frequency.
To make this work self-contained, we also provide a comprehensive summary of quantum statistics in Appendix~\ref{appa}, derive the generalized Ehrenfest theorem in Appendix~\ref{appb}, and prove the Robertson--Schr\"{o}dinger uncertainty relations in Appendix~\ref{appc}.

\section{Mandelstam--Tamm Uncertainty Relation}
\label{sec:2}

Mandelstam and Tamm~\cite{Mandelstam1945}   derived the so-called time--energy
uncertainty relation based on the Robertson uncertainty relation~\eqref{eq:43} for two Hermitian operators $\hat{A}=\hat{A}^\dagger$ and $\hat{B}=\hat{B}^\dagger$, in which they have set $\hat{B}=\hat{H}$ 
\begin{equation}
\Delta\hat{A}\cdot\Delta\hat{H}\geq\frac{1}{2}\left|\left\langle \left[\hat{A},\hat{H}\right]\right\rangle \right|
\label{eq:55}
\end{equation}
where $\Delta$ indicates the \emph{{standard deviation}} of the given quantum observable (cf.~\eqref{eq:SD}) and the quantum Hamiltonian $\hat{H}$ is related to the total energy of the quantum system.
After the substitution of $\imath\hbar\frac{d}{dt} \langle \hat{A} \rangle $
for the expectation of the commutator $ \langle [\hat{A},\hat{H} ] \rangle $
based on the generalized Ehrenfest theorem~\eqref{eq:Ehrenfest},
we obtain
\begin{equation}
\Delta\hat{A}\cdot\Delta\hat{H}\geq\frac{\hbar}{2}\left|\frac{d}{dt}\left\langle \hat{A}\right\rangle \right|
.
\end{equation}
Now, the  identification of $\Delta E\equiv\Delta\hat{H}$ and 
\begin{equation}
\Delta T\equiv\frac{\Delta\hat{A}}{\left|\frac{d}{dt}\left\langle \hat{A}\right\rangle \right|}
\label{eq:Tamm}
\end{equation}
results in
\begin{equation}
\Delta E\cdot\Delta T\geq\frac{\hbar}{2}
\label{eq:MTU}.
\end{equation}
The Mandelstam--Tamm uncertainty relation has been also generalized for mixed\linebreak states \cite{Uhlmann1992,Hornedal2022,Bagchi2023}; however, the \emph{physical interpretation} of $\Delta T$ has not been scrutinized in detail.

Before we focus on the physical interpretation of $\Delta T$, it would be useful to recall several fundamental theorems on the quantum statistics of the total energy $\hat{H}$ and its expectation value $\langle\hat{H}\rangle$, variance $\textrm{Var}(\hat{H})$, and standard deviation $\Delta \hat{H}$.

\section{Conservation of Total Energy and Its Quantum Statistics}
\label{sec:3}

\begin{Theorem}
\label{thm:1}
For a closed quantum system, the expectation value of the total energy $\langle \hat{H} \rangle $
does not change in time
\begin{equation}
\frac{d}{dt}\left\langle \hat{H}\right\rangle =0
.
\end{equation}
Furthermore, the expectation values for individual energy eigenvectors do not change in time, which implies that the variance $\textrm{Var}(\hat{H})$ and standard deviation $\Delta \hat{H}$ of energy also do not change in time.
\end{Theorem}
\begin{proof}
The application of the generalized Ehrenfest theorem~\eqref{eq:Ehrenfest}
immediately gives
\begin{equation}
\frac{d}{dt}\left\langle \hat{H}\right\rangle =\frac{1}{\imath\hbar}\left\langle \left[\hat{H},\hat{H}\right]\right\rangle =0
.
\end{equation}
Alternatively, one can start from the general basis-independent solution
of the Schr\"{o}dinger equation
\begin{equation}
|\Psi(t)\rangle=e^{-\frac{\imath}{\hbar}\hat{H}t}|\Psi(0)\rangle
\label{eq:sol}
\end{equation}
and express it in the energy basis as
\begin{align}
|\Psi(t)\rangle & =\sum_{i=1}^{n}\alpha_{i}\left(0\right)e^{-\frac{\imath}{\hbar}E_{i}t}|E_{i}\rangle
\end{align}
where $\left\{ |E_{i}\rangle\right\}_{i=1}^n $ is the complete set of \emph{energy eigenvectors} that span the Hilbert space~$\mathcal{H}$ of the quantum system,
$\left\{ E_{i}\right\}_{i=1}^n $ is the set of corresponding \emph{energy eigenvalues}, and the initial energy quantum probability amplitudes are $\alpha_{i}\left(0\right)=\langle E_{i}|\Psi\left(0\right)\rangle$.

Since the spectral decomposition of the Hamiltonian is
\begin{equation}
\hat{H}=\sum_{k=1}^{n}E_{k}|E_{k}\rangle\langle E_{k}|
\end{equation}
for any time $t$, we have 
\begin{align}
\langle\Psi\left(t\right)|\hat{H}|\Psi(t)\rangle & =\sum_{j=1}^{n}\alpha_{j}^{*}\left(0\right)e^{+\frac{\imath}{\hbar}E_{j}t}\langle E_{j}|\left(\sum_{k=1}^{n}E_{k}|E_{k}\rangle\langle E_{k}|\right)\sum_{i=1}^{n}\alpha_{i}\left(0\right)e^{-\frac{\imath}{\hbar}E_{i}t}|E_{i}\rangle\nonumber \\
 & =\sum_{i=1}^{n}\sum_{k=1}^{n}\sum_{j=1}^{n}\alpha_{i}\left(0\right)\alpha_{j}^{*}\left(0\right)E_{k}e^{\frac{\imath}{\hbar}\left(E_{j}-E_{i}\right)t}\langle E_{j}|E_{k}\rangle\langle E_{k}|E_{i}\rangle\nonumber \\
 & =\sum_{i=1}^{n}\sum_{k=1}^{n}\sum_{j=1}^{n}\alpha_{i}\left(0\right)\alpha_{j}^{*}\left(0\right)E_{k}e^{\frac{\imath}{\hbar}\left(E_{j}-E_{i}\right)t}\delta_{jk}\delta_{ki}\nonumber \\
 & =\sum_{i=1}^{n}\left|\alpha_{i}\left(0\right)\right|^{2}E_{i}=\langle\Psi\left(0\right)|\hat{H}|\Psi(0)\rangle
.
\end{align}
Each quantum probability amplitude for the corresponding energy eigenstate
\begin{equation}
	\alpha_{i}\left(t\right)=\alpha_{i}\left(0\right)e^{-\frac{\imath}{\hbar}E_{i}t}
	\label{eq:amplitude}
\end{equation}
oscillates in the Hilbert space $\mathcal{H}$ with angular frequency $\omega_{i}=\frac{E_{i}}{\hbar}$,
whereas the quantum probability $\left|\alpha_{i}\left(t\right)\right|^{2}=\left|\alpha_{i}\left(0\right)\right|^{2}$
remains constant, $\frac{d}{dt}\left|\alpha_{i}\left(t\right)\right|^{2}=0$.
\end{proof}

Theorem~\ref{thm:1} shows that for a closed quantum system, characterized by quantum Hamiltonian~$\hat{H}$ and initial quantum state $\Psi\left(0\right)\rangle$, the quantum statistical quantities of energy are stable system properties that do not change in time.

\begin{Definition}
For a quantum state vector $|\Psi\rangle$ that is expressed in some given basis~$\{|\psi_{i}\rangle\}_{i=1}^n$ of $n$-dimensional Hilbert space~$\mathcal{H}$, the basis-dependent $\ell_1$ \emph{norm of coherence}~$\mathcal{C}$ and \emph{predictability}~$\mathcal{P}$ are defined~\cite{Baumgratz2014,Qureshi2019,Qureshi2021,Peled2020} as
\begin{align}
\mathcal{C} & =\frac{1}{n-1}\sum_{i\neq j}\left|\alpha_{i}\right|\left|\alpha_{j}\right| ,\\
\mathcal{P} & =\sqrt{1-\mathcal{C}^{2}} ,
\end{align}
where $\alpha_{i}=\langle\psi_{i}|\Psi\rangle$ is the quantum probability
amplitude corresponding to each unit vector $|\psi_{i}\rangle$ in
the complete basis set resolving the unit operator $\hat{I} =\sum_{i=1}^n|\psi_{i}\rangle\langle\psi_{i}|$, namely
\begin{equation}
|\Psi\rangle=\hat{I}|\Psi\rangle=\sum_{i=1}^n|\psi_{i}\rangle\langle\psi_{i}||\Psi\rangle=\sum_{i=1}^n\alpha_{i}|\psi_{i}\rangle
.
\end{equation}
\end{Definition}

\begin{Theorem}
\label{thm:1b}
For a closed quantum system, the $\ell_1$ norm of energy coherence~$\mathcal{C}$ and energy predictability~$\mathcal{P}$ are constant in time.
\end{Theorem}
\begin{proof}
Using the explicit time dynamics~\eqref{eq:amplitude} for each energy eigenstate, we have
\begin{align}
\mathcal{C}(t) & =\frac{1}{n-1}\sum_{i\neq j}\left|\alpha_{i}(t)\right|\left|\alpha_{j}(t)\right|=\frac{1}{n-1}\sum_{i\neq j}\left|\alpha_{i}(0)\right|\left|e^{-\frac{\imath}{\hbar}E_{i}t}\right|\left|\alpha_{j}(0)\right|\left|e^{-\frac{\imath}{\hbar}E_{j}t}\right| \nonumber \\
 & =\frac{1}{n-1}\sum_{i\neq j}\left|\alpha_{i}(0)\right|\left|\alpha_{j}(0)\right|=\mathcal{C}(0) ,\\
\mathcal{P}(t) & =\sqrt{1-\mathcal{C}^{2}(t)} = \sqrt{1-\mathcal{C}^{2}(0)} = \mathcal{P}(0)
.
\end{align}
\end{proof}

Theorem~\ref{thm:1b} suggests that because the $\ell_1$ norm of energy coherence~$\mathcal{C}$ is a time-invariant physical property, it can be used as a convenient classification of quantum systems with a given quantum Hamiltonian $\hat{H}$ and initial quantum state $|\Psi(0)\rangle $.

\section{Minimal Quantum Toy Model}
\label{sec:4}

Consider a spin-$\frac{1}{2}$ quantum particle (qubit) in a uniform
static magnetic field $\vec{B}$ that is aligned in the $z$-direction~\cite{Georgiev2021}.
The quantum Hamiltonian of the system is
\begin{equation}
\hat{H}=\frac{1}{2}\hbar\omega\,\hat{\sigma}_{z}=\frac{1}{2}\hbar\omega\left(\begin{array}{cc}
1 & 0\\
0 & -1
\end{array}\right)
\label{eq:Hamiltonian}
\end{equation}
where $\hat{\sigma}_{z}$ is the Pauli spin matrix aligned in the $z$-direction.

\subsection{Quantum Dynamics of Energy States}

The energy eigenstates of the quantum system are the eigenstates of
the Hamiltonian~\eqref{eq:Hamiltonian}, namely, $|\uparrow_{z}\rangle$,
$|\downarrow_{z}\rangle$ with corresponding eigenvalues $+\frac{1}{2}\hbar\omega,-\frac{1}{2}\hbar\omega$.

\begin{figure}[H]
\includegraphics[width=\textwidth]{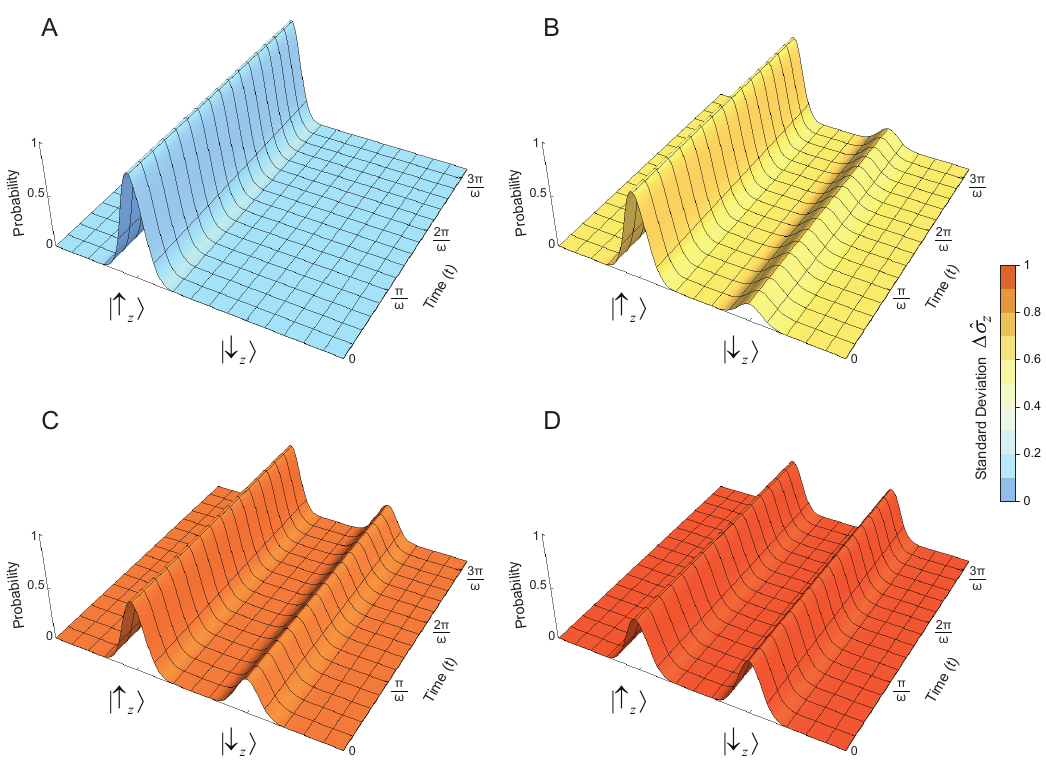}
\caption{\label{fig:1} The quantum dynamics of energy eigenstates is trivial; that is,
the expectation value for each energy eigenvector does not change
in time. This implies that the standard deviation of energy is constant
and fixed by the choice of initial quantum state. Based on the quantum
Hamiltonian~\eqref{eq:Hamiltonian}, the eigenvectors of $\hat{\sigma}_{z}$
play the role of energy eigenvectors. The quantum superposition of
the energy quantum probability amplitudes of the initial state $|\Psi(0)\rangle$
given in~\eqref{eq:psi0} is varied as follows:
(\textbf{A})~$\alpha_{1}=1$, $\alpha_{2}=0$, $\mathcal{C}=0$,
(\textbf{B})~$\alpha_{1}=\sqrt{5/6}$, $\alpha_{2}=\sqrt{1/6}$, $\mathcal{C}\approx 0.745$,
(\textbf{C})~$\alpha_{1}=\sqrt{2/3}$, $\alpha_{2}=\sqrt{1/3}$, $\mathcal{C}\approx 0.943$,
(\textbf{D})~$\alpha_{1}=\sqrt{1/2}$, $\alpha_{2}=\sqrt{1/2}$, $\mathcal{C}=1$.}
\end{figure}

The \emph{matrix exponential} of the Hamiltonian is transformed into an ordinary
exponential of the corresponding energy eigenvalues when acting on
the energy eigenstates
\begin{align}
e^{-\frac{\imath}{\hbar}\hat{H}t}|\uparrow_{z}\rangle & 
=e^{-\frac{1}{2}\imath\omega t}|\uparrow_{z}\rangle ,\\
e^{-\frac{\imath}{\hbar}\hat{H}t}|\downarrow_{z}\rangle & 
=e^{+\frac{1}{2}\imath\omega t}|\downarrow_{z}\rangle
.
\end{align}
Therefore, it is useful to express the initial state $|\Psi(0)\rangle$
in the energy eigenbasis
\begin{equation}
|\Psi(0)\rangle=\alpha_{1}|\uparrow_{z}\rangle+\alpha_{2}|\downarrow_{z}\rangle
\label{eq:psi0}
\end{equation}
with normalization condition
\begin{equation}
\sum_{i}\left|\alpha_{i}\right|^{2}=\left|\alpha_{1}\right|^{2}+\left|\alpha_{2}\right|^{2}=1
.
\end{equation}
The general solution of the Schr\"{o}dinger Equation~\eqref{eq:sol} in the energy basis becomes
\begin{equation}
|\Psi(t)\rangle=\alpha_{1}e^{-\frac{1}{2}\imath\omega t}|\uparrow_{z}\rangle+\alpha_{2}e^{+\frac{1}{2}\imath\omega t}|\downarrow_{z}\rangle
\label{eq:psit}
.
\end{equation}
The \emph{{expectation value}} of the energy is independent of time (Figure~\ref{fig:1})
\begin{equation}
\left\langle \hat{H}\right\rangle =\frac{1}{2}\hbar\omega\left\langle \hat{\sigma}_{z}\right\rangle =\frac{1}{2}\hbar\omega\left(\left|\alpha_{1}\right|^{2}-\left|\alpha_{2}\right|^{2}\right)
.
\end{equation}

Similarly, the \emph{{variance}} of the energy is independent of time
\begin{align}
\textrm{Var}\left(\hat{H}\right) & =\left\langle \hat{H}^{2}\right\rangle -\left\langle \hat{H}\right\rangle ^{2}=\frac{1}{4}\hbar^{2}\omega^{2}\textrm{Var}\left(\hat{\sigma}_{z}\right) ,\\
\textrm{Var}\left(\hat{\sigma}_{z}\right) & =\left\langle \hat{\sigma}_{z}^{2}\right\rangle -\left\langle \hat{\sigma}_{z}\right\rangle ^{2}=\left[1^2 -\left(\left|\alpha_{1}\right|^{2}-\left|\alpha_{2}\right|^{2}\right)^{2}\right]\nonumber \\
 & =\left[\left(\left|\alpha_{1}\right|^{2}+\left|\alpha_{2}\right|^{2}\right)^{2}-\left(\left|\alpha_{1}\right|^{2}-\left|\alpha_{2}\right|^{2}\right)^{2}\right]\nonumber \\
 & =4\left|\alpha_{1}\right|^{2}\left|\alpha_{2}\right|^{2}
,
\end{align}
and the \emph{{standard deviation}} of the energy is independent of time
\begin{align}
\Delta\hat{H} & =\frac{1}{2}\hbar\omega\Delta\hat{\sigma}_{z} = \frac{1}{2}\hbar\omega \mathcal{C}
=\hbar\omega\left|\alpha_{1}\right|\left|\alpha_{2}\right|\equiv\Delta E \label{eq:delE} ,\\
\Delta\hat{\sigma}_{z} & =2\left|\alpha_{1}\right|\left|\alpha_{2}\right|
.
\end{align}
The $\ell_1$ norm of energy coherence~$\mathcal{C}$ and predictability~$\mathcal{P}$~\cite{Peled2020,Qureshi2019,Qureshi2021}
are related to the standard deviation and expectation value of $\hat{\sigma}_{z}$, respectively
\begin{align}
\mathcal{C} & =\Delta\hat{\sigma}_{z}=2\left|\alpha_{1}\right|\left|\alpha_{2}\right| ,\\
\mathcal{P} & =\left|\left\langle \hat{\sigma}_{z}\right\rangle \right|=\left|\left|\alpha_{1}\right|^{2}-\left|\alpha_{2}\right|^{2}\right|
,
\end{align}
and satisfy the \emph{{complementarity relation}}
\begin{equation}
\mathcal{P}^{2}+\mathcal{C}^{2} = 1
.
\end{equation}

\subsection{Quantum Dynamics of Eigenstates of the Clock Observable}

From the generalized Ehrenfest theorem~\eqref{eq:Ehrenfest}, it is clear that any quantum observable that commutes with the Hamiltonian $\hat{H}$ will be static and cannot be used to measure time. Consequently, to construct a clock, one needs to consider a quantum observable that does not commute with $\hat{H}$. Since in the minimal quantum toy model the Hamiltonian is a scaled version of $\hat{\sigma}_{z}$, it is interesting to check the behavior of a mutually unbiased quantum observable, i.e., $\hat{\sigma}_{x}$ or $\hat{\sigma}_{y}$. Without loss of generality, here, we choose $\hat{\sigma}_{x}$ as a \emph{clock observable}.

\begin{figure}[H]
\includegraphics[width=\textwidth]{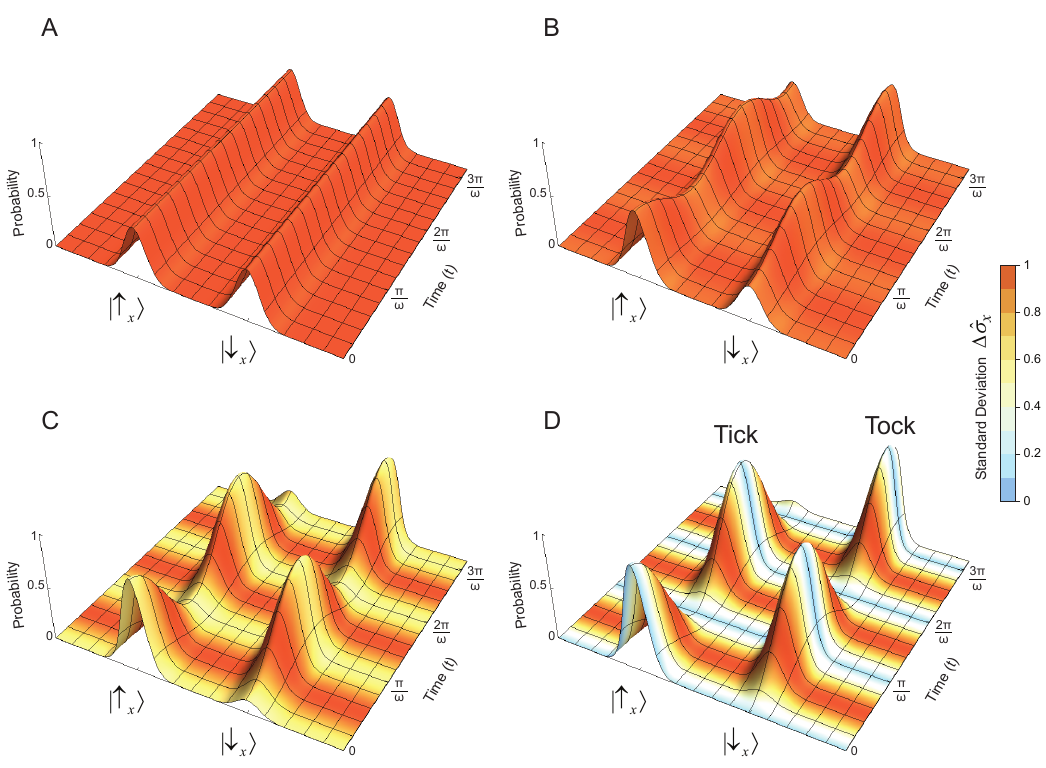}
\caption{\label{fig:2} Quantum dynamics of the eigenvectors of $\hat{\sigma}_{x}$
is not trivial as long as the initial state of the quantum system
is not an energy eigenstate. Each eigenvector of $\hat{\sigma}_{x}$
is a quantum superposition of energy eigenvectors according to~\eqref{eq:75}
and~\eqref{eq:76}. The quantum superposition of the energy quantum
probability amplitudes of the initial state $|\Psi(0)\rangle$ given
in~\eqref{eq:psi0} is varied as follows:
(\textbf{A})~$\alpha_{1}=1$, $\alpha_{2}=0$, $\mathcal{C}=0$,
(\textbf{B})~$\alpha_{1}=\sqrt{39/40}$, $\alpha_{2}=\sqrt{1/40}$, $\mathcal{C}\approx 0.312$,
(\textbf{C})~$\alpha_{1}=\sqrt{5/6}$, $\alpha_{2}=\sqrt{1/6}$, $\mathcal{C}\approx 0.745$,
(\textbf{D})~$\alpha_{1}=\sqrt{1/2}$, $\alpha_{2}=\sqrt{1/2}$, $\mathcal{C}=1$.
The labels ``Tick'' and ``Tock'' indicate how the maxima or minima of the dynamic expectation value $\langle\hat{\sigma}_{x}\rangle$ can be used for the engineering of a physical clock.}
\end{figure}

The expectation values of eigenvectors of $\hat{\sigma}_{x}$ change in time because
the eigenvectors of $\hat{\sigma}_{x}$ are quantum superpositions
of eigenvectors of $\hat{\sigma}_{z}$ as follows 
\begin{align}
|\uparrow_{x}\rangle & =\frac{1}{\sqrt{2}}\left(|\uparrow_{z}\rangle+|\downarrow_{z}\rangle\right)\label{eq:75} ,\\
|\downarrow_{x}\rangle & =\frac{1}{\sqrt{2}}\left(|\uparrow_{z}\rangle-|\downarrow_{z}\rangle\right)\label{eq:76} .
\end{align}
To change the basis from $\left\{ |\uparrow_{z}\rangle,|\downarrow_{z}\rangle\right\} $
to $\left\{ |\uparrow_{x}\rangle,|\downarrow_{x}\rangle\right\} $,
we can add or subtract~\eqref{eq:75} and~\eqref{eq:76} in order
to obtain
\begin{align}
|\uparrow_{z}\rangle & =\frac{1}{\sqrt{2}}\left(|\uparrow_{x}\rangle+|\downarrow_{x}\rangle\right) ,\\
|\uparrow_{z}\rangle & =\frac{1}{\sqrt{2}}\left(|\uparrow_{x}\rangle-|\downarrow_{x}\rangle\right) .
\end{align}
The substitution in~\eqref{eq:psit} gives
\begin{align}
|\Psi(t)\rangle 
& =\alpha_{1}e^{-\frac{1}{2}\imath\omega t}\frac{1}{\sqrt{2}}\left(|\uparrow_{x}\rangle+|\downarrow_{x}\rangle\right)+\alpha_{2}e^{+\frac{1}{2}\imath\omega t}\frac{1}{\sqrt{2}}\left(|\uparrow_{x}\rangle-|\downarrow_{x}\rangle\right)\nonumber \\
 & =\frac{1}{\sqrt{2}}\left(\alpha_{1}e^{-\frac{1}{2}\imath\omega t}+\alpha_{2}e^{+\frac{1}{2}\imath\omega t}\right)|\uparrow_{x}\rangle+\frac{1}{\sqrt{2}}\left(\alpha_{1}e^{-\frac{1}{2}\imath\omega t}-\alpha_{2}e^{+\frac{1}{2}\imath\omega t}\right)|\downarrow_{x}\rangle
.
\end{align}
The expectation values of the projectors $|\uparrow_{x}\rangle\langle\uparrow_{x}|$
and $|\downarrow_{x}\rangle\langle\downarrow_{x}|$ are
\begin{align}
\langle\Psi(t)|\uparrow_{x}\rangle\langle\uparrow_{x}|\Psi(t)\rangle & =\frac{1}{2}\left|\alpha_{1}e^{-\frac{1}{2}\imath\omega t}+\alpha_{2}e^{+\frac{1}{2}\imath\omega t}\right|^{2}
=\frac{1}{2}\left|\alpha_{1}+\alpha_{2}e^{\imath\omega t}\right|^{2}\nonumber \\
 & =\frac{1}{2}+\textrm{Re}\left(\alpha_{1}\alpha_{2}^{*}\right)\cos\left(\omega t\right)+\textrm{Im}\left(\alpha_{1}\alpha_{2}^{*}\right)\sin\left(\omega t\right) \label{eq:xup},\\
\langle\Psi(t)|\downarrow_{x}\rangle\langle\downarrow_{x}|\Psi(t)\rangle & =\frac{1}{2}\left|\alpha_{1}e^{-\frac{1}{2}\imath\omega t}-\alpha_{2}e^{+\frac{1}{2}\imath\omega t}\right|^{2}
=\frac{1}{2}\left|\alpha_{1}-\alpha_{2}e^{\imath\omega t}\right|^{2} \nonumber \\
 & =\frac{1}{2}-\textrm{Re}\left(\alpha_{1}\alpha_{2}^{*}\right)\cos\left(\omega t\right)-\textrm{Im}\left(\alpha_{1}\alpha_{2}^{*}\right)\sin\left(\omega t\right) \label{eq:xdown}
.
\end{align}
{The quantum dynamics of the quantum observables given by the projectors $|\uparrow_{x}\rangle\langle\uparrow_{x}|$
and $|\downarrow_{x}\rangle\langle\downarrow_{x}|$ onto the eigenvectors of $\hat{\sigma}_{x}$
exhibits oscillations with an angular frequency $\omega$ whenever the initial state of the quantum system
is not an energy eigenstate (Figure~\ref{fig:2}).} 

\vspace{-6pt}

Since the eigenvalues for $\left\{ |\uparrow_{x}\rangle,|\downarrow_{x}\rangle\right\} $
are $\left\{ +1,-1\right\} $, respectively, we subtract~\eqref{eq:xdown} from~\eqref{eq:xup} in order to obtain the expectation value of $\hat{\sigma}_{x}$ and its temporal dynamics
\begin{align}
\left\langle \hat{\sigma}_{x}\right\rangle  
 & =2\left[\textrm{Re}\left(\alpha_{1}\alpha_{2}^{*}\right)\cos\left(\omega t\right)+\textrm{Im}\left(\alpha_{1}\alpha_{2}^{*}\right)\sin\left(\omega t\right)\right], \\
\frac{d}{dt}\left\langle \hat{\sigma}_{x}\right\rangle &=-2\omega\left[\textrm{Re}\left(\alpha_{1}\alpha_{2}^{*}\right)\sin\left(\omega t\right)-\textrm{Im}\left(\alpha_{1}\alpha_{2}^{*}\right)\cos\left(\omega t\right)\right]
\label{eq:dtsx},
\end{align}
For the variance and standard deviation of $\hat{\sigma}_{x}$, we have
\begin{align}
\textrm{Var}\left(\hat{\sigma}_{x}\right) 
& =\left\langle \hat{\sigma}_{x}^{2}\right\rangle -\left\langle \hat{\sigma}_{x}\right\rangle ^{2}
= 1 -\left\langle \hat{\sigma}_{x}\right\rangle ^{2}
\nonumber \\
 & =1^2 -\left(\frac{1}{2}\left|\alpha_{1}e^{-\frac{1}{2}\imath\omega t}+\alpha_{2}e^{+\frac{1}{2}\imath\omega t}\right|^{2}-\frac{1}{2}\left|\alpha_{1}e^{-\frac{1}{2}\imath\omega t}-\alpha_{2}e^{+\frac{1}{2}\imath\omega t}\right|^{2}\right)^{2}\nonumber \\
 & =\left|\alpha_{1}e^{-\frac{1}{2}\imath\omega t}+\alpha_{2}e^{+\frac{1}{2}\imath\omega t}\right|^{2}\left|\alpha_{1}e^{-\frac{1}{2}\imath\omega t}-\alpha_{2}e^{+\frac{1}{2}\imath\omega t}\right|^{2} \nonumber\\
 & = \left|\alpha_{1}^{2}e^{-\imath\omega t}-\alpha_{2}^{2}e^{+\imath\omega t}\right|^{2} 
=\left|\alpha_{1}^{2}-\alpha_{2}^{2}e^{\imath 2\omega t}\right|^{2} 
= 1-2\left|\alpha_{1}\right|^{2}\left|\alpha_{2}\right|^{2}-2\textrm{Re}\left[\left(\alpha_{1}\alpha_{2}^{*}\right)^{2}e^{-\imath2\omega t}\right] \nonumber\\
 & = \left|\alpha_{1}\right|^{4}+\left|\alpha_{2}\right|^{4}-2\textrm{Re}\left[\left(\alpha_{1}\alpha_{2}^{*}\right)^{2}\right]\cos\left(2\omega t\right)-2\textrm{Im}\left[\left(\alpha_{1}\alpha_{2}^{*}\right)^{2}\right]\sin\left(2\omega t\right) ,\\
\Delta\hat{\sigma}_{x}  
& = \left|\alpha_{1}^{2}e^{-\imath\omega t}-\alpha_{2}^{2}e^{+\imath\omega t}\right|
\label{eq:delsx}
.
\end{align}

\subsection{Physical Meaning of Mandelstam--Tamm ``Time Uncertainty''}

\begin{figure}[H]
\includegraphics[width=\textwidth]{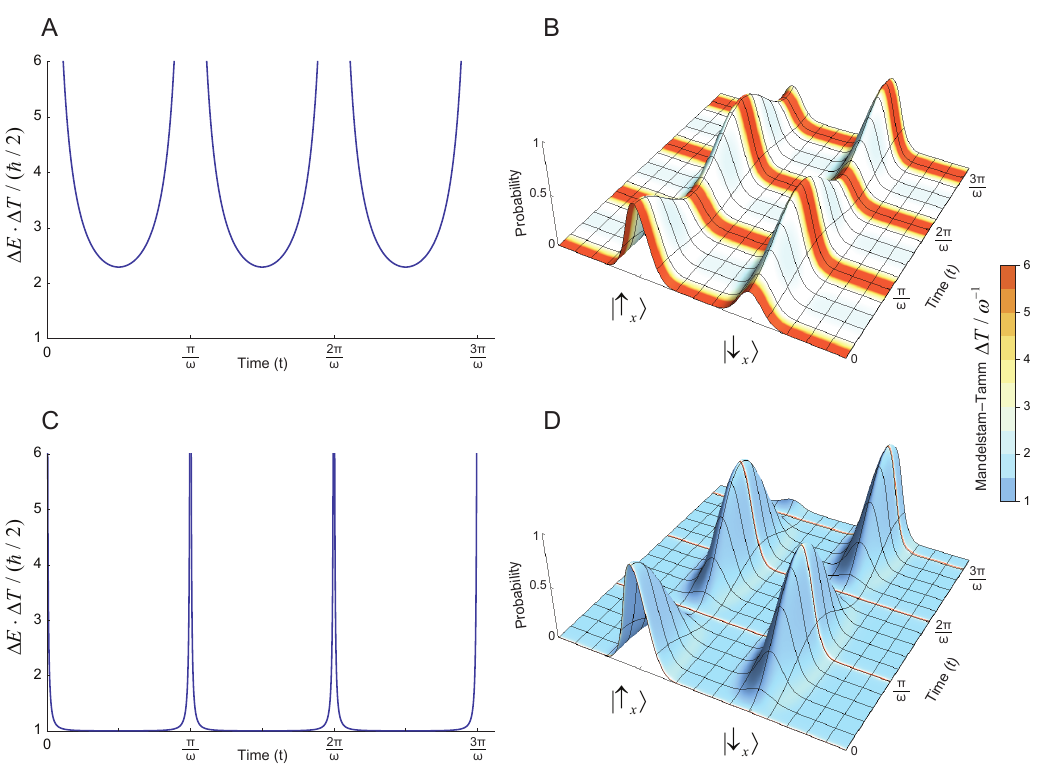}
\caption{\label{fig:3} Time dynamics of the product $\Delta E\cdot\Delta T (t)$ (measured
in units of $\hbar/2$) or $\Delta T (t)$ (measured in units of $\omega^{-1}$)
for the initial state $|\Psi(0)\rangle$ given in~\eqref{eq:psi0} with
\mbox{(\textbf{A},\textbf{B})}~$\alpha_{1}=\sqrt{19/20}$, $\alpha_{2}=\sqrt{1/20}$, $\mathcal{C}\approx 0.436$, or 
\mbox{(\textbf{C},\textbf{D})}~$\alpha_{1}=\sqrt{11/20}$, $\alpha_{2}=\sqrt{9/20}$, $\mathcal{C}\approx 0.995$.}
\end{figure}

Mandelstam and Tamm~\cite{Mandelstam1945} defined ``time uncertainty'' to be
\begin{equation}
\Delta T\equiv\frac{\Delta\hat{\sigma}_{x}}{\left|\frac{d}{dt}\left\langle \hat{\sigma}_{x}\right\rangle \right|}
\label{eq:MTsx}
.
\end{equation}
The substitution of~\eqref{eq:dtsx} and~\eqref{eq:delsx} in~\eqref{eq:MTsx} gives
\begin{align}
\Delta T=\frac{\left|\alpha_{1}^{2}e^{-\imath\omega t}-\alpha_{2}^{2}e^{+\imath\omega t}\right|}{2\omega\left|\textrm{Re}\left(\alpha_{1}\alpha_{2}^{*}\right)\sin\left(\omega t\right)-\textrm{Im}\left(\alpha_{1}\alpha_{2}^{*}\right)\cos\left(\omega t\right)\right|}
\label{eq:MTsx2}
.
\end{align}
After combining~\eqref{eq:delE} and~\eqref{eq:MTsx2}, the time--energy uncertainty relation \eqref{eq:MTU} becomes
\begin{equation}
\Delta E\cdot\Delta T = \frac{\hbar\omega\left|\alpha_{1}\right|\left|\alpha_{2}\right|\left|\alpha_{1}^{2}e^{-\imath\omega t}-\alpha_{2}^{2}e^{+\imath\omega t}\right|}{2\omega\left|\textrm{Re}\left(\alpha_{1}\alpha_{2}^{*}\right)\sin\left(\omega t\right)-\textrm{Im}\left(\alpha_{1}\alpha_{2}^{*}\right)\cos\left(\omega t\right)\right|}\geq\frac{\hbar}{2}
.
\end{equation}
In general, the Mandelstam--Tamm quantity $\Delta T (t)$~\eqref{eq:MTsx2}
exhibits a dynamic dependence on time $t$. Because the standard deviation
of energy $\Delta E$ is constant, the Mandelstam--Tamm
product $\Delta E\cdot\Delta T(t)$ also exhibits
a dynamic dependence on time $t$ inherited from~$\Delta T (t)$. From the dynamic plots shown in
Figure~\ref{fig:3}, it can be observed that $\Delta T(t)$ does not meaningfully
correspond either to physical \emph{time}~$t$ or to the \emph{angular frequency}~$\omega$
of oscillation of the expectation value $\left\langle \hat{\sigma}_{x}\right\rangle $.
In particular, $\Delta T(t)\to\infty$ at the instances of local maxima
or minima of $\left\langle \hat{\sigma}_{x}\right\rangle $ at which
$\frac{d}{dt}\left\langle \hat{\sigma}_{x}\right\rangle =0$. It can
be stated that the angular frequency $\omega$ of oscillation of the
expectation value $\left\langle \hat{\sigma}_{x}\right\rangle $ is
independent of the initial state $|\Psi(0)\rangle$; however, the
observable amplitude of oscillation measured by the difference
$\left\langle \hat{\sigma}_{x}\right\rangle _{\max}-\left\langle \hat{\sigma}_{x}\right\rangle _{\min}$
tends to zero as $\mathcal{C}\to 0$ (Figure~\ref{fig:2}A,B).

\vspace{-6pt}

Only for the very special case with $\mathcal{C}= 1$ shown in Figure~\ref{fig:2}D,
we arrive at Mandelstam--Tamm quantity $\Delta T$
that is independent of time due to the precise cancellation of the numerator
and the denominator 
\begin{equation}
\Delta T=\frac{|\sin\left(\omega t\right)|}{\left|\frac{d}{dt}\cos\left(\omega t\right)\right|}=\frac{1}{\omega}=\frac{t}{2\pi}
,
\end{equation}
which results in minimal Mandelstam--Tamm uncertainty:
\begin{equation}
\Delta T\cdot\Delta E=\frac{1}{\omega}\cdot\frac{\hbar\omega}{2}=\frac{\hbar}{2}
.
\end{equation}

Mandelstam and Tamm~\cite{Mandelstam1945} have claimed that $\Delta T$ \emph{{represents the amount of time it takes the expectation value of any quantum observable}} $\hat{A}$ \emph{{to change by one standard deviation}} $\Delta \hat{A}$. This claim is incorrectly justified in textbooks~\cite{Griffiths2018} by rewriting~\eqref{eq:Tamm} in the form
\begin{equation}
	\Delta\hat{A}= |\frac{d}{dt} \langle \hat{A} \rangle| ~\Delta T
\end{equation}
and then interpreting $\Delta\hat{A}$ as distance, $|\frac{d}{dt} \langle \hat{A} \rangle|$ as speed, and $\Delta T$ as time.
Here, we have shown that in general, $\Delta T (t)$ is a dynamic function of time~$t$ and can become arbitrarily large in the vicinity of local maxima or minima of the dynamic expectation value $\langle \hat{A}\rangle (t)$, where the instantaneous rate of change vanishes $|\frac{d}{dt} \langle \hat{A} \rangle| (t) = 0$ (Figure~\ref{fig:3}).
In fact, by correctly considering that all three quantities in~\eqref{eq:Tamm} are time-dependent, namely
\begin{equation}
	\Delta\hat{A} (t)= |\frac{d}{dt} \langle \hat{A} \rangle| (t) ~\Delta T (t) ,
\end{equation}
the correct description of $\Delta T (t)$ is \emph{{an instantaneous ``hypothetical time duration'' that would have been taken by the quantum observable moving with a constant speed equal to the instantaneous rate of change}} $|\frac{d}{dt} \langle \hat{A} \rangle| (t)$ \emph{{until a distance equal to the instantaneous standard deviation}} $\Delta\hat{A} (t)$ \emph{{would have been traversed}}.
Because the rate of change $|\frac{d}{dt} \langle \hat{A} \rangle| (t)$ accelerates or decelerates, $\Delta T (t)$~is an \emph{{instantaneous quantity}} that does not in itself disclose how good the quantum system is in measuring the physical passage of time.
This is to be contrasted with $\Delta E$, which is a \emph{{constant quantity}} that discloses how good the quantum system is at measuring the physical passage of time.

\subsection{Clock Engineering with the Einstein--Planck Relation}

The engineering of a physical clock based on the time dynamics of
$\left\langle \hat{\sigma}_{x}\right\rangle $ requires
a coherent quantum superposition of at least two energy eigenvectors
with distinct energy eigenvalues, $E_{1}\neq E_{2}$, in the initial
state $|\Psi(0)\rangle$. Then, oscillations of $\left\langle \hat{\sigma}_{x}\right\rangle $
can be observed with an angular frequency determined by the Einstein--Planck relation
\begin{equation}
\omega=\frac{\left|E_{1}-E_{2}\right|}{\hbar}
\label{eq:Einstein}
.
\end{equation}
Although the time~$t$ is a parameter and does not have an associated
quantum operator in nonrelativistic quantum mechanics, it is possible
to measure time using the time dynamics of the expectation value of some quantum
observable, such as $\left\langle \hat{\sigma}_{x}\right\rangle $,
whose eigenvectors form a mutually unbiased basis with regard
to the energy eigenbasis, which is given by the eigenvectors of $\hat{\sigma}_{z}$
in the minimal quantum toy model. For the overwhelming majority of
initial states $|\Psi(0)\rangle$ of the quantum system, the Mandelstam--Tamm
quantity $\Delta T$ will dynamically change in time $t$ and will
be poorly suited to act as a measure of time.
For example, let us define that
at the local maxima of $\left\langle \hat{\sigma}_{x}\right\rangle $,
the physical clock has generated a ``tick'',
whereas at the local minima of $\left\langle \hat{\sigma}_{x}\right\rangle $,
the physical clock has generated a ``tock''.
The time that passes between the consecutive ``tick'' at time $t_1$ and ``tock'' at time $t_2$ (Figure~\ref{fig:2}D) is exactly the time period
\begin{equation}
	\delta t= |t_2 - t_1 | =\frac{\pi}{\omega}
.
\end{equation}
At the maxima or minima of the expectation  value of the clock observable $\left\langle \hat{\sigma}_{x}\right\rangle $, we could say that we know exactly \emph{what time it is} measured in units of $\delta t$, i.e., $1\delta t, 2\delta t, 3\delta t, \ldots $
\mbox{In between} the maxima or minima of~$\left\langle \hat{\sigma}_{x}\right\rangle $, we may not know how much time has passed since the last ``tick'' or ``tock'' of the clock because the dynamic quantum state is in a coherent quantum superposition of the eigenvectors of~$\hat{\sigma}_{x}$.
If~we~set $\delta E = \left|E_{1}-E_{2}\right|$, we can rewrite the Einstein--Planck relation~\eqref{eq:Einstein} as
\begin{equation}
	\delta E \cdot \delta t=\frac{h}{2}
\label{eq:new}
.
\end{equation}
Equation~\eqref{eq:new} is a strict relationship between scalars, $\delta$ denotes \emph{{difference}} instead of \emph{{standard deviation}}, and the expression contains $h$ instead of $\hbar$ when compared to~\eqref{eq:MTU}.
This shows that the time--energy relationship in nonrelativistic quantum mechanics is due to the Einstein--Planck relation~\eqref{eq:Einstein}  and not due to the Heisenberg uncertainty principle generalized through the Robertson--Schr\"{o}dinger inequality~\eqref{eq:44}.
This is further corroborated by the fact that, at the minima or maxima of $\left\langle \hat{\sigma}_{x}\right\rangle $, the Mandelstam--Tamm
quantity $\Delta T$ will tend to infinity and would wrongly predict
that the standard deviation of time is infinite, while in fact we know exactly how much time has passed in units of $\delta t$.
This is why the Mandelstam--Tamm quantity $\Delta T$ is not a meaningful representation of time, and
the Mandelstam--Tamm uncertainty is misleadingly labeled as ``time--energy
uncertainty'' relationship.

In essence, the generalized Ehrenfest theorem~\eqref{eq:Ehrenfest}
correctly predicts that the time dynamics of quantum observables is only
possible if the initial quantum state is not an energy eigenstate.
The construction of physical clocks that are capable of measuring
time, however, is aided by the Einstein--Planck relation~\eqref{eq:Einstein}
regardless of divergent Mandelstam--Tamm quantity $\Delta T$ near
the local maxima of the measured quantum observable at which the physical clock ``ticks''.

\section{Margolus--Levitin Quantum Speed Limit}
\label{sec:5}

Margolus and Levitin~\cite{Margolus1998} have recognized that  time $t$ indeed has no
variance in nonrelativistic quantum mechanics, and they have proposed 
measuring $\tau_{\perp}$, which is \emph{{the time it takes for}} $|\Psi(0)\rangle$
\emph{{to evolve into an orthogonal state}}. Furthermore, they have stated
not only an inequality based on the standard deviation of the energy
\begin{equation}
\tau_{\perp}\geq\frac{h}{4\Delta\hat{H}}
=\frac{\pi\hbar}{2\Delta\hat{H}}\label{eq:levi1}
\end{equation}
but also, in their own words, they have formulated an alternative ``somewhat surprizing'' result based
on the expectation value of the energy
\begin{equation}
\tau_{\perp}\geq\frac{h}{4\left\langle \hat{H}\right\rangle }
=\frac{\pi\hbar}{2\left\langle \hat{H}\right\rangle }\label{eq:levi2}
.
\end{equation}
We will first show that in fact~\eqref{eq:levi2} is incorrect as
stated in nonrelativistic quantum mechanics because adding a constant
energy offset to the Hamiltonian will change the expectation value
of the Hamiltonian without changing the quantum dynamics.

\begin{Theorem}
\label{thm:2}
(Constant offset to the Hamiltonian) Adding a constant offset $E_{0}$
in the Hamiltonian $\hat{H}=\hat{H}_{0}+E_{0}\hat{I}$ of a closed system has no observable effect on the quantum dynamics.
\end{Theorem}
\begin{proof}
Because the identity operator commutes with every operator, the application
of the Baker--Campbell--Hausdorff formula~\cite{Campbell1897,Baker1905,Hausdorff1906,Dynkin1947} gives the following result for the general basis-independent solution to the Schr\"{o}dinger equation
\begin{equation}
|\Psi(t)\rangle=e^{-\frac{\imath}{\hbar}\left(\hat{H}_{0}+E_{0}\hat{I}\right)t}|\Psi(0)\rangle=e^{-\frac{\imath}{\hbar}\hat{H}_{0}t}e^{-\frac{\imath}{\hbar}E_{0}\hat{I}t}|\Psi(0)\rangle=e^{-\frac{\imath}{\hbar}\hat{H}_{0}t}e^{-\frac{\imath}{\hbar}E_{0}t}|\Psi(0)\rangle
,
\end{equation}
which means that the energy offset $E_{0}$ simply adds a global time-evolving
pure phase $e^{-\frac{\imath}{\hbar}E_{0}t}$ to every state. This
global pure phase has no observable effects on measured quantum probabilities
since it has a unit modulus, $\left|e^{-\frac{\imath}{\hbar}E_{0}t}\right|^{2}=1$.
Thus, the freedom to add a constant energy offset to the Hamiltonian of a closed system is granted by the $U(1)$~\emph{symmetry} of the global (overall) phase of the quantum mechanical wavefunction~$\Psi$, according to which the global (overall) phase of $\Psi$ can never be measured, whereas only relative phases of $\Psi$ can be measured experimentally.
\end{proof}

Theorem~\ref{thm:2} always makes it  possible to offset the Hamiltonian so that it has an  expectation value of zero, $\langle \hat{H}\rangle = 0$, which forces~\eqref{eq:levi2} to produce incorrect $\tau_{\perp} \geq \infty$. In fact, for the quantum toy model shown in Figure~\ref{fig:2}D, it is exactly the case that $\langle \hat{H}\rangle = 0$ with computed $\tau_{\perp} \geq \infty$ from~\eqref{eq:levi2}, instead of the correct $\tau_{\perp} = \frac{\pi}{\omega}$ seen on the plot. This shows that the Margolus--Levitin formulation~\eqref{eq:levi2} based on the expectation value of energy $\langle \hat{H}\rangle$ is only meaningful in the context of their particular choice to set the zero energy level to be the minimal energy eigenvalue, $E_{\min}=0$. In our single-qubit toy example, the zero energy level is set to be the arithmetic mean (average) of the two energy eigenvalues, $\frac{1}{2} (E_{\max}+E_{\min}) = 0$, so that the quantum Hamiltonian has a symmetric matrix representation. To use~\eqref{eq:levi2}, one needs to properly offset the Hamiltonian~\eqref{eq:Hamiltonian} with the addition of $\frac{1}{2} \hbar\omega \hat{I}$.

Theorem~\ref{thm:2} explains why the most general studies of quantum dynamics use statistical properties such as variance or standard deviation that are \emph{invariant} with respect to offset shifts in the expectation value of energy.
In fact,~\eqref{eq:levi1} always holds regardless of the energy expectation value. For the case $\mathcal{C}=1$ shown in Figure~\ref{fig:2}D, plugging $\Delta \hat{H} = \frac{1}{2}\hbar \omega$ in~\eqref{eq:levi1} results in the formally correct inequality $\frac{\pi}{\omega}\geq \frac{\pi}{\omega}$.
However, we will show that for all cases in which $\mathcal{C}<1$, and  hence $|\alpha_1|^2\neq|\alpha_2|^2$, the quantum system never evolves into an orthogonal state, meaning that $\tau_{\perp} = \infty$ and~\eqref{eq:levi1} reduces to the trivial inequality $\infty \geq \frac{\hbar\pi}{2\Delta\hat{H}}$, which is always true.

To compute $\tau_{\perp}$ for the minimal quantum toy model, we need the minimal value of $t$ at which $\left|\langle\Psi(0)|\Psi(t)\rangle\right|=0$. The inner product formed by~\eqref{eq:psi0} and~\eqref{eq:psit} is
\begin{align}
\left|\langle\Psi(0)|\Psi(t)\rangle\right| 
& =\left|\left(\alpha_{1}^{*}\langle\uparrow_{z}|+\alpha_{2}^{*}\langle\downarrow_{z}|\right)
(\alpha_{1}e^{-\frac{1}{2}\imath\omega t}|\uparrow_{z}\rangle+\alpha_{2}e^{+\frac{1}{2}\imath\omega t}|\downarrow_{z}\rangle )\right|\nonumber \\
 & =\left|\left|\alpha_{1}\right|^{2}e^{-\frac{1}{2}\imath\omega t}+\left|\alpha_{2}\right|^{2}e^{+\frac{1}{2}\imath\omega t}\right| 
  = \left|e^{+\frac{1}{2}\imath\omega t}\right|\left|\left|\alpha_{1}\right|^{2}e^{-\frac{1}{2}\imath\omega t}+\left|\alpha_{2}\right|^{2}e^{+\frac{1}{2}\imath\omega t}\right| \nonumber \\
 & = \left|\left|\alpha_{1}\right|^{2}+\left|\alpha_{2}\right|^{2}e^{\imath\omega t}\right| 
  \nonumber\\
& \geq\left|\left|\alpha_{1}\right|^{2}-\left|\alpha_{2}\right|^{2}\right|=\mathcal{P}=\sqrt{1-\mathcal{C}^{2}}
\label{eq:fail}
.
\end{align}
From~\eqref{eq:fail}, it follows that for $\mathcal{C}<1$, i.e., $|\alpha_1|^2\neq|\alpha_2|^2$, the predictability of the energy eigenstate is strictly non-negative $\mathcal{P}>0$, implying that the quantum system starting from an initial quantum state $|\Psi(0)\rangle$ never evolves into an orthogonal state, i.e.,\linebreak $\tau_{\perp} = \infty$.
Thus, the Margolus--Levitin formulation~\eqref{eq:levi1} is trivially satisfied and has no physical content for almost all initial quantum states (for which $\mathcal{C}<1$), while the only special case when~\eqref{eq:levi1} actually works is for $\mathcal{C}=1$, i.e., $|\alpha_1|^2=|\alpha_2|^2$, leading to the strict equality\linebreak $\tau_{\perp} 4 \Delta\hat{H} = \tau_{\perp} 2 |E_{\max} - E_{\min}| = h$. Taking into consideration that $\tau_{\perp} = \frac{1}{2\nu}$ is the time period from maximum to minimum, rather than between two maxima of the oscillation of $\langle\hat{\sigma}_x\rangle$, we conclude that~\eqref{eq:levi1} is just the Einstein--Planck relation $|E_{\max} - E_{\min}| = h\nu = \hbar\omega$ in disguise.

It is noteworthy that   the calculation in~\eqref{eq:fail} can be straightforwardly generalized for the quantum state vector $|\Psi(t)\rangle$ in $n$-dimensional Hilbert space. If we arrange the initial energy amplitudes of $|\Psi(0)\rangle$ in decreasing order $\left|\alpha_{1}\right|\geq\left|\alpha_{2}\right|\geq\ldots\geq\left|\alpha_{n}\right|$, we obtain
\begin{align}
\left|\langle\Psi(0)|\Psi(t)\rangle\right| & =\left|\left|\alpha_{1}\right|^{2}e^{-\imath\frac{E_{1}}{\hbar}t}+\left|\alpha_{2}\right|^{2}e^{-\imath\frac{E_{2}}{\hbar}t}+\ldots+\left|\alpha_{n}\right|^{2}e^{-\imath\frac{E_{n}}{\hbar}t}\right|\nonumber \\
 & \geq\left|\alpha_{1}\right|^{2}-\left(\left|\alpha_{2}\right|^{2}+\ldots+\left|\alpha_{n}\right|^{2}\right)
,
\end{align}
which, when combined with the normalization condition $\sum_n \left|\alpha_{n}\right|^2 = 1$, gives $\tau_{\perp} = \infty$ for all cases with $\left|\alpha_{1}\right|^2 > \frac{1}{2}$. The minimal $\tau_{\perp} = h/(2|E_{\max} - E_{\min}|)$ is obtained when the initial energy amplitudes of the energy eigenvectors $|E_{\max}\rangle$ and $|E_{\min}\rangle$, with maximal and minimal eigenvalues,  respectively, have  a modulus of $1/\sqrt{2}$.

{Several authors~\cite {Giovannetti2003,Levitin2009,Deffner2017} have previously discussed a unified \emph{{quantum speed limit}}~(QSL) given by the maximum of the Mandelstam--Tamm and Margolus--Levitin lower bounds on the ``time uncertainty''
\begin{equation}
	\tau_\textrm{QSL} = \max\left\{ \frac{\pi\hbar}{2 \Delta \hat{H}} , \frac{\pi\hbar}{2 \langle \hat{H}\rangle} \right \}
	= \frac{h}{4 \times \min\left\{\Delta \hat{H} , \langle \hat{H}\rangle \right\}}
	\label{eq:sql}
\end{equation}
where the energy expectation value $\langle \hat{H}\rangle$ is computed with the particular offset to the Hamiltonian for which $E_{\min}=0$.
Here, we have shown that $\tau_\textrm{QSL}$ is generally neither Mandelstam--Tamm $\Delta T$ nor Margolus--Levitin $\tau_{\perp}$, since for the submaximal positive $\ell_1$~norm of energy coherence, $0<\mathcal{C}<1$, in the single-qubit toy model, both $\Delta T$ and $\tau_{\perp}$ could be infinite, whereas $\tau_\textrm{QSL}$ given by~\eqref{eq:sql} is always finite.}

\section{Conclusions}
\label{sec:6}

In this work, we have investigated the purported role of Heisenberg's uncertainty principle~\cite{Heisenberg1927} in establishing time--energy uncertainty relations in nonrelativistic quantum mechanics. Using a single qubit inside an external magnetic field as a minimal quantum toy model, we have investigated two different time--energy inequalities formulated by Mandelstam and Tamm~\cite{Mandelstam1945} and by Margolus and Levitin~\cite{Margolus1998}.
We have shown that both of these time--energy inequalities are plagued by infinities and do not meaningfully represent  the concept of time for general initial states $|\Psi(0)\rangle$ with the submaximal $\ell_1$ norm of energy coherence  $\mathcal{C}<1$. Importantly, for the special case of initial quantum state $|\Psi(0)\rangle$ with the maximal $\ell_1$ norm of energy coherence  $\mathcal{C}=1$, both Mandelstam--Tamm and  Margolus--Levitin inequalities reduce to the Einstein--Planck relation \mbox{$|E_{\max} - E_{\min}| = h \nu =\hbar\omega$}, thereby acquiring concrete physical meaning. Thus, the shortest duration of time measurable using some quantum observable acting as a clock is not due to Heisenberg's uncertainty principle  but follows directly from the Einstein--Planck relation. This explains why for probing physical processes at shorter timescales, particle physicists need to use larger particle accelerators generating higher particle energies.

We have also elaborated on the fact that adding a constant energy offset to the Hamiltonian does not affect the quantum dynamics of closed systems. This explains why statistical properties such as variance~$\textrm{Var}(\hat{H})$, standard deviation~$\Delta \hat{H}$, and the $\ell_1$ norm of energy coherence~$\mathcal{C}$, all of which are invariant with respect to shifts in the energy expectation value~$\langle \hat{H} \rangle$, are well suited for the  classification of quantum systems wih regard to the magnitude of observed quantum dynamical changes, namely, when $\textrm{Var}(\hat{H})$, $\Delta \hat{H}$, and $\mathcal{C}$ approach zero, the observed amplitude of oscillation of dynamic quantum observables also approaches zero.

\vspace{6pt} 


\funding{This research was funded by Cosmogenics Inc., Los Angeles, CA, USA.}

\dataavailability{Data sharing is not applicable to this theoretical research article as no datasets were generated or analyzed during the current study.}

\acknowledgments{The author wishes to acknowledge helpful discussions with Mani L. Bhaumik (University of California, Los Angeles).}

\conflictsofinterest{The author declares no conflicts of interest.} 

\appendixtitles{yes} 
\appendixstart
\appendix

\section{Quantum Statistics}
\label{appa}

\begin{Definition}
(Expectation value) The expectation value of any quantum observable represented by a Hermitian operator
$\hat{X}=\hat{X}^\dagger$ is a real number written in angle brackets as $\langle\hat{X}\rangle$.
For a discrete spectrum of eigenvalues
$x_{i}$ of $\hat{X}$, we have
\begin{equation}
\langle\hat{X}\rangle=\sum_{i}x_{i}p_{i}
\end{equation}
where $p_{i}\in[0,1]$ is the probability of obtaining the
measurement outcome $x_{i}$.

For a continuous probability density distribution $p(x)=\psi^{*}\left(x\right)\psi\left(x\right)$,
we use the integration
\begin{equation}
\langle\hat{X}\rangle=\int_{-\infty}^{\infty}x\,p(x)dx
\label{eq:11}
.
\end{equation}
The expectation value is functionally dependent on the quantum
state vector $|\psi\rangle$, namely
\begin{equation}
\langle\hat{X}\rangle\equiv\langle\psi|\hat{X}|\psi\rangle=\langle\hat{X}\rangle_{\psi}
.
\end{equation}
To emphasize the dependence on the quantum state $|\psi\rangle$, the expectation value could be written as~$\langle\hat{X}\rangle_{\psi}$~\cite{Georgiev2022}.
However, if it is clear from the context
 which $|\psi\rangle$ we perform the calculation for, it is notationally
simpler to write just $\langle\hat{X}\rangle$.
\end{Definition}

\begin{Definition}
(Variance) The variance of a quantum observable $\hat{X}=\hat{X}^\dagger$ is a real non-negative number denoted
as $\textrm{Var}(\hat{X})$. 
For a discrete spectrum of eigenvalues $x_{i}$ of $\hat{X}$, we have
\begin{equation}
\textrm{Var}(\hat{X})=\left\langle \left(\hat{X}-\left\langle \hat{X}\right\rangle \right)^{2}\right\rangle 
= \left\langle \hat{X}^{2}\right\rangle -\left\langle \hat{X}\right\rangle ^{2}
\label{eq:13}
.
\end{equation}
For a continuous probability density distribution $p(x)$, we use integration
\begin{equation}
\textrm{Var}(\hat{X})=\int_{-\infty}^{\infty}x^{2}p(x)dx-\left(\int_{-\infty}^{\infty}x\,p(x)dx\right)^{2}
\label{eq:14}
.
\end{equation}
\end{Definition}

\begin{Definition}
(Standard deviation) The standard deviation is the square root of the variance
\begin{equation}
\Delta\hat{X}=\sqrt{\textrm{Var}(\hat{X})}
\label{eq:SD}
.
\end{equation}
\end{Definition}

\section{Generalized Ehrenfest Theorem}
\label{appb}

\begin{Theorem}
(Generalized Ehrenfest theorem) The time dynamics of the expectation value
$\langle\hat{A}\rangle$ of any time-independent quantum observable
$\hat{A}$ (for which $\frac{d}{dt}\hat{A}=0$) is given by
\begin{equation}
\frac{d}{dt}\left\langle \hat{A}\right\rangle =\frac{1}{\imath\hbar}\left\langle \left[\hat{A},\hat{H}\right]\right\rangle \label{eq:Ehrenfest}
\end{equation}
where the commutator $\left[\hat{A},\hat{H}\right]=\hat{A}\hat{H}-\hat{H}\hat{A}$
is with respect to the quantum Hamiltonian~$\hat{H}$.
\end{Theorem}
\begin{proof}
The theorem follows directly from the product rule for differentiation
and the Schr\"{o}dinger Equation~\eqref{eq:1} (cf.~\cite{Georgiev2019})
\begin{equation}
\frac{d}{dt}\left\langle \hat{A}\right\rangle =\frac{d}{dt}\left\langle \Psi|\hat{A}|\Psi\right\rangle =\left(\frac{d}{dt}\langle\Psi|\right)\hat{A}|\Psi\rangle+\langle\Psi|\left(\frac{d}{dt}\hat{A}\right)|\Psi\rangle+\langle\Psi|\hat{A}\left(\frac{d}{dt}|\Psi\rangle\right)\label{eq:32}
.
\end{equation}
From the Schr\"{o}dinger Equation~\eqref{eq:1}, we have
\begin{align}
\frac{d}{dt}|\Psi\rangle & =\frac{1}{\imath\hbar}\hat{H}|\Psi\rangle\label{eq:33} ,\\
\frac{d}{dt}\langle\Psi| & =-\frac{1}{\imath\hbar}\langle\Psi|\hat{H}\label{eq:34} ,
\end{align}
where we used the Hermiticity of the Hamiltonian $\hat{H}=\hat{H}^{\dagger}$.
After the substitution of~\eqref{eq:33} and~\eqref{eq:34} together with
$\frac{d}{dt}\hat{A}=0$ inside~\eqref{eq:32}, we obtain
\begin{align}
\frac{d}{dt}\left\langle \Psi|\hat{A}|\Psi\right\rangle 
 & =\frac{1}{\imath\hbar}\langle\Psi|\left(\hat{A}\hat{H}-\hat{H}\hat{A}\right)|\Psi\rangle
.
\end{align}
\end{proof}

\section{Robertson--Schr\"{o}dinger Uncertainty Relation}
\label{appc}

\begin{Theorem}
(Robertson uncertainty relation) For any two quantum observables given
by Hermitian operators $\hat{A}=\hat{A}^{\dagger}$ and $\hat{B}=\hat{B}^{\dagger}$,
for which $\hat{A}|\Psi\rangle$ is in the domain of $\hat{B}$ and
$\hat{B}|\Psi\rangle$ is in the domain of $\hat{A}$~\cite{Davidson1965},
the following inequality holds~\cite{Robertson1929}
\begin{equation}
\Delta\hat{A}\cdot\Delta\hat{B}\geq\frac{1}{2}\left|\left\langle \left[\hat{A},\hat{B}\right]\right\rangle \right|\label{eq:43}
.
\end{equation}
\end{Theorem}
\begin{proof}
The Robertson uncertainty relation is a special case of the more general Schr\"{o}dinger
uncertainty relation~\eqref{eq:44} proven below.
\end{proof}

\begin{Theorem}
(Schr\"{o}dinger uncertainty relation) For any two quantum observables
given by Hermitian operators $\hat{A}$ and $\hat{B}$, for which
$\hat{A}|\Psi\rangle$ is in the domain of~$\hat{B}$ and $\hat{B}|\Psi\rangle$
is in the domain of~$\hat{A}$~\cite{Davidson1965}, 
the following inequality holds~\cite{Schrodinger1930,S1,S2}
\begin{equation}
\Delta\hat{A}\cdot\Delta\hat{B}\geq\sqrt{\left|\frac{1}{2}\left\langle \hat{A}\hat{B}+\hat{B}\hat{A}\right\rangle -\left\langle \hat{A}\right\rangle \left\langle \hat{B}\right\rangle \right|^{2}+\left|\frac{1}{2}\left\langle \left[\hat{A},\hat{B}\right]\right\rangle \right|^{2}}
\label{eq:44}
.
\end{equation}
\end{Theorem}
\begin{proof}
The quantum observables are represented by Hermitian operators $\hat{A}=\hat{A}^{\dagger}$
and $\hat{B}=\hat{B}^{\dagger}$. Therefore, the variances can be
written as inner products of vectors
\begin{align}
\textrm{Var}(\hat{A})
& =\left(\Delta\hat{A}\right)^{2} 
=\left\langle \Psi\right|\left(\hat{A}^{\dagger}-\left\langle \hat{A}^{\dagger}\right\rangle \right)
\left(\hat{A}-\left\langle \hat{A}\right\rangle \right)\left|\Psi\right\rangle \label{eq:45}
,\\
\textrm{Var}(\hat{B})
& =\left(\Delta\hat{B}\right)^{2}
=\left\langle \Psi\right|\left(\hat{B}^{\dagger}-\left\langle \hat{B}^{\dagger}\right\rangle \right)
\left(\hat{B}-\left\langle \hat{B}\right\rangle \right)\left|\Psi\right\rangle \label{eq:46}
.
\end{align}
Now, we can apply the Cauchy--Schwarz inequality
\begin{equation}
\langle a|a\rangle\langle b|b\rangle\geq\left|\langle a|b\rangle\right|^{2}
\end{equation}
with
\begin{align}
|a\rangle & = \left(\hat{A}-\left\langle \hat{A}\right\rangle \right)|\Psi\rangle ,\\
|b\rangle & = \left(\hat{B}-\left\langle \hat{B}\right\rangle \right)|\Psi\rangle ,
\end{align}
to obtain
\begin{equation}
\left(\Delta\hat{A}\right)^{2}\left(\Delta\hat{B}\right)^{2}\geq\left|\langle\Psi|\left(\hat{A}-\left\langle \hat{A}\right\rangle \right)\left(\hat{B}-\left\langle \hat{B}\right\rangle \right)|\Psi\rangle\right|^{2}
.
\end{equation}
Since $z=\langle a|b\rangle$ is a complex number, we have $z^{*}=\langle b|a\rangle$
and
\begin{equation}
\left|z\right|^{2}=z^{*}z=\left[\textrm{Re}\left(z\right)\right]^{2}+\left[\textrm{Im}\left(z\right)\right]^{2}=\left[\frac{z+z^{*}}{2}\right]^{2}+\left[\frac{z-z^{*}}{2\imath}\right]^{2}\label{eq:51}
.
\end{equation}
We also have
\begin{align}
z & =\langle a|b\rangle=\left\langle \hat{A}\hat{B}\right\rangle -\left\langle \hat{A}\right\rangle \left\langle \hat{B}\right\rangle \label{eq:52} ,\\
z^{*} & =\langle b|a\rangle=\left\langle \hat{B}\hat{A}\right\rangle -\left\langle \hat{A}\right\rangle \left\langle \hat{B}\right\rangle \label{eq:53}
.
\end{align}
The substitution of~\eqref{eq:52} and~\eqref{eq:53} into~\eqref{eq:51}
gives
\begin{equation}
\left(\Delta\hat{A}\right)^{2}\left(\Delta\hat{B}\right)^{2}\geq\left[\frac{\left\langle \hat{A}\hat{B}+\hat{B}\hat{A}\right\rangle -2\left\langle \hat{A}\right\rangle \left\langle \hat{B}\right\rangle }{2}\right]^{2}+\left[\frac{\left\langle \hat{A}\hat{B}\right\rangle -\left\langle \hat{B}\hat{A}\right\rangle }{2\imath}\right]^{2}
,
\end{equation}
which, after   introducing  the commutator $\left[\hat{A},\hat{B}\right]$
and taking the square root on both sides, leads to~\eqref{eq:44}.
\end{proof}

\begin{adjustwidth}{-\extralength}{0cm}
\reftitle{References}

\PublishersNote{}
\end{adjustwidth}

\end{document}